\documentclass[draft]{agujournal2019}
\usepackage{url} 
\usepackage{lineno}
\usepackage[inline]{trackchanges} 
\usepackage{soul}
\usepackage{natbib}

\draftfalse

\journalname{JGR: Space Physics}

\begin{document}

%
%


\title{First Results of Calculation of the Dayside Magnetopause Reconnection Rate with TRACERS}

%
%




\authors{D. E. da Silva$^{1,2}$, L. J. Chen$^{1}$, B. Burkholder$^{1,2}$, C. J. Martin$^{3}$, B. Michotte de Welle$^{1}$, I. DesJardin$^{1,4}$,  S. A. Fuselier$^{5,6}$, K. J. Trattner$^{7}$, S. M. Petrinec$^{8}$, G. Cucho-Padin$^{1,4}$, D. Sur$^{1,4}$, N. Y. Buzulukova$^{9}$, J. Bonnell$^{10}$, J. S. Halekas$^{11}$, D. Miles$^{11}$, S. Henderson$^{11}$, J. Dorelli$^{1}$}

\affiliation{1}{Heliophysics Sciences Division, NASA Goddard Spaceflight Center, Greenbelt, MD}
\affiliation{2}{Goddard Planetary Heliophysics Institute, University of Maryland Baltimore County, Baltimore, MD}
\affiliation{3}{Institute of Space and Atmospheric Studies, University of Saskatchewan, Saskatoon, SK, Canada}
\affiliation{4}{The Catholic University of America, Washington, DC}
\affiliation{5}{Southwest Research Institute, San Antonio, TX}
\affiliation{6}{University of Texas at San Antonio, San Antonio, TX}
\affiliation{7}{Laboratory for Atmospheric and Space Physics, University of Colorado, Boulder, CO}
\affiliation{8}{Lockheed Martin Advanced Technology Center, Palo Alto, CA}
\affiliation{9} {Department of Astronomy, University of Maryland, College Park, MD}
\affiliation{10}{Space Sciences Laboratory, University of California, Berkeley, CA}
\affiliation{11}{University of Iowa, Iowa City, IA}





\correspondingauthor{Daniel E. da Silva}{Daniel.E.daSilva@nasa.gov}



\begin{keypoints}
\item First results of estimation of the dayside magnetopause reconnection rate from cusp ion-energy dispersion using TRACERS are presented.
\item A correlation of $r^2=0.62$ ($N=5$) is found between the reconnection rate and SuperDARN cross polar cap potential.
\item An existing algorithm developed to automatically identify dispersion events is adapted to TRACERS data and validated.
\end{keypoints}

%
%

%
%


\begin{abstract}
Quantifying the rate of reconnection remains a key challenge in the study of geomagnetic storms. In this study, we present the first results from the new Tandem Reconnection and Cusp Electrodynamics Reconnaissance Satellites (TRACERS) mission to calculate the reconnection rate from dispersed ion populations measured in the cusp, using previously developed methodology. We study ion dispersion events in two storms. During the first storm, which occurred on September 30, 2025, 21 dispersion events were identified in TRACERS 2 ion data as a combined result of geospace dynamics and the spacecraft cusp encounters. Throughout this time when ion dispersion was detected by TRACERS , the reconnection rate is calculated and is shown to track reasonably well with another measure of geomagnetic activity, the Cross Polar Cap Potential (CPCP) as obtained by the Super Dual Auroral Radar Network (SuperDARN). The dispersions are correlated with the smoothed CPCP, and a correlation of $r^2=0.62$ was found using the dispersions from the onset and main phase of the storm. The correlation disappears when including dispersions from the recovery period. The distribution of reconnection rates throughout this storm lies generally between 0.1 - 1.5 mV/m, slightly lower than previous reconnection rates calculated with DMSP, but consistent with in-situ observations made with Polar. In this work, we also apply a previously developed automatic identification algorithm for ion-energy cusp dispersion to TRACERS, with the intent that it may serve as a tool for future statistical analysis as the data volume becomes too large to manually review.

\end{abstract}


%
%

\section{Introduction}

Geomagnetic storms at Earth are a primary cause of space weather effects, in which the solar wind interacts with Earth's magnetic field through a process known as magnetic reconnection. Magnetic reconnection can occur at multiple places throughout the magnetosphere, with the classic scenario occurring at the dayside magnetopause where a southward-pointing magnetic field from the solar wind meets the Earth's dayside northward-pointing magnetic field \citep{dungey1961interplanetary,kivelson1996introduction}. At locations where the magnetic shear is large, merging alters the magnetic topology. Plasma entrained on these newly reconnected field lines are then accelerated to higher energies at the expense of a reduction in the magnetic field tension. The nominal and local degree to which magnetic reconnection occurs, as measured through the reconnection electric field at the magnetopause, is known as the reconnection rate. This reconnection rate is closely related to the velocity of the reconnected plasma around the Earth and the speed that the fields lines are moving, setting the speed of the Dungey cycle \citep{dungey1961interplanetary}.


The magnetospheric cusps are funnel-shaped regions in Earth's magnetic field in both hemispheres. Because of the cusp's connection to the dayside magnetic topology and the ability for particles to flow freely along the cusp field line into low Earth orbit, the cusp has been a critical region used to study the global magnetic reconnection processes \citep{lockwood1994low,smith1996earth,wing2001double,wing2005cusp,trattner2005reconnection,trattner1999spatial,trattner2003cusp}.

Many missions have been used to study the cusp, at both low Earth orbit (LEO; typically circular orbits under 2000 km altitude) and High Earth Orbit (HEO; nominally with an apogee above the geostationary altitude, 6.6 $R_E$). LEO missions studying the cusp include the Defense Meteorological Satellite Program (DMSP) \citep{redmon2017new}, which spanned multiple decades of cusp observations with continuously replaced satellites. HEO missions have also collected data on the cusp, including the Cluster mission \citep{escoubet2001introduction}. An advantage of observing the cusp in LEO over HEO is that the revisit cadence of cusp crossings for LEO satellites is much higher than for HEO satellites; the orbital period for Cluster is 54 hours, while the orbital period for DMSP is 101 minutes \citep{da2025calculation}. This short revisit time makes it possible to have multiple cusp crossings throughout the main phase of a storm with LEO satellites.

With revisit times significantly less than a storm duration, it is desirable to develop methods to estimate the local or global dayside magnetopause reconnection rate and quantify reconnection over a storm. A method of calculating the dayside magnetopause reconnection rate from dispersed ions in the cusps was first proposed in \citet{lockwood1992variation} and further developed in  \citet{da2025calculation}, \citet{burkholder2025simulated}, and \citet{burkholder2026day}.

In this work, we discuss the first results on the calculation of the dayside magnetopause reconnection rate through observations of dispersed particle precipitation in the cusps, using the new Tandem Reconnection and Cusp Electrodynamics Reconnaissance Satellites mission (TRACERS; \citealp{miles2025tandem,petrinec2025tandem}). The TRACERS mission design is centered on studying magnetic reconnection at the cusps, with two satellites orbiting in LEO at an altitude of 590 km (periods of 96.68 minutes). Two key instruments aboard each TRACERS satellite for this study are the Analyzer for Cusp Ions (ACI; \citealp{fuselier2025analyzer}) and the Analyzer for Cusp Electrons (ACE; \citealp{halekas2025tracers}). The two TRACERS satellites are in the same time-delayed orbit, with a minimum temporal separation of 10 seconds.

\begin{figure}
    \centering
    \includegraphics[width=1\linewidth]{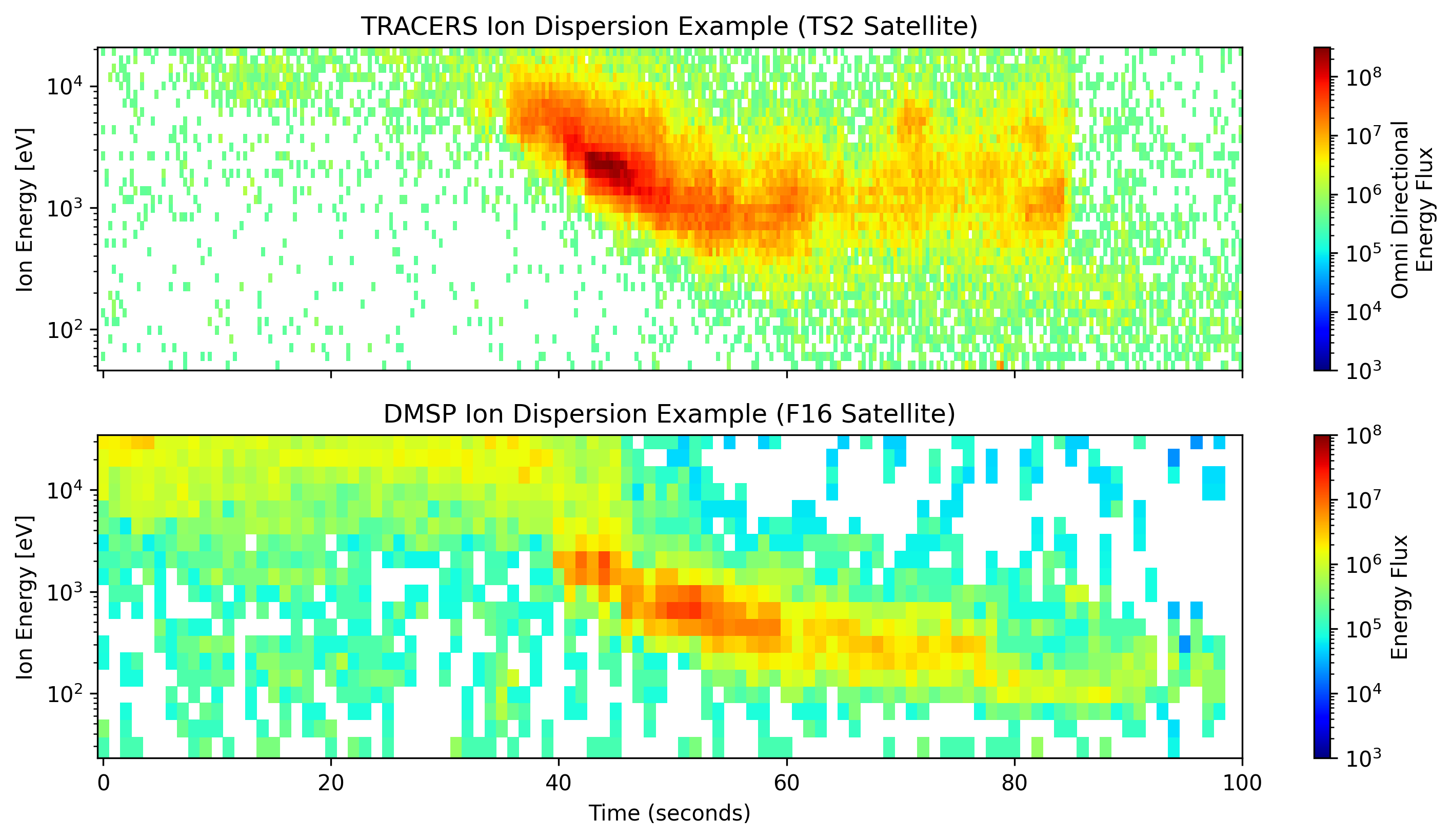}
    \caption{Illustration of the ability to resolve ion-energy dispersion structures in low Earth orbit between the new capability (on TRACERS) and the predecessor capability (on DMSP). Note: these are two different events. The time resolution of an spectrogram energy sweep with DMSP/SSJ (the ion/electron instrument) is 1 second, while the time resolution of a full 4$\pi$ sr velocity distribution with TRACERS/ACI (the ion instrument) is 312 ms. The TRACERS event starts at November 15, 2025, 21:42:00 UTC, while the DMSP event starts at December 21, 2015, 04:09:50 UTC.}
    \label{fig:dmsp_vs_tracers}
\end{figure}

The instrumentation aboard TRACERS represents a major improvement over DMSP for monitoring the cusps from LEO. For reference, a comparison between TRACERS and DMSP for two dispersed ion events is shown in Figure \ref{fig:dmsp_vs_tracers}. The top panel shows an event measured by TRACERS, and the bottom panel shows a similar event measured by DMSP. The 312 ms time resolution of TRACERS ion measurements, compared to 1 second time resolution for DMSP, allows micro-structures within the larger dispersion structure to be better discerned. 

Furthermore, DMSP's SSJ instrument, which measures ions and electrons, was limited to a 90-degree FOV that was susceptible to irregular coverage of the pitch angle distribution \citep{redmon2017new,da2025calculation}. The ACI instrument on TRACERS does not have this limitation and, in fact, provides full pitch-angle distributions if desired.

In this work, we study two storms during the first six months of the TRACERS mission: the Sept 30, 2025, storm and the January 19, 2026 storm. We adapt the methodology from \citep{da2022automatic} to automatically select dispersion events and identify the lower energy cutoff, and then apply the methodology for calculating the reconnection rate to TRACERS \citep{da2025calculation,burkholder2025simulated,burkholder2026day}.  The improved instrumentation in TRACERS/ACI, compared with the previously used DMSP, enables unprecedented precision in calculating the reconnection rate.

The outline of this article is as follows. In Section \ref{sec:data_and_models}, we describe the data and models used in this study. In Section \ref{sec:methodology}, we review the automatic identification methodology, the hand validation, and the methodology for calculating the dayside magnetopause reconnection rate from dispersed ion signatures. Section \ref{sec:storms} presents our results for the two geomagnetic storms, including plots of reconnection rates during each storm, a correlation analysis with CPCP, and the overall histogram of observed reconnection rates. Finally, Section \ref{sec:conclusion} discusses the initial results and avenues for future studies of the reconnection rate using TRACERS.

\section{Data and Models} \label{sec:data_and_models}

TRACERS is a two-satellite mission that launched in July, 2025, consisting of nominal satellites TRACERS-1 and TRACERS-2. 
Due to operational constraints, our dataset for the September 30, 2025, storm relies exclusively on TRACERS-2 measurements. In contrast, the January 19th, 2026, event includes observations from both spacecraft, albeit with unequal data coverage.

During the September 30th, 2025 storm, TRACERS-2 was collecting high-resolution data in the northern cusp, while during the January 19th, 2026 storm, the operations had switched to collecting measurements in the southern cusp. The ion instrument on TRACERS, ACI, collects data over 47 energy channels between 8 eV and 20,000 eV at 312 ms time resolution \citep{fuselier2025analyzer}. However, due to issues at low energies, we omit the bottom 10 energy channels from our analysis.  The electron instrument, ACE, is not contaminated at low energies and covers 49 energy channels between 20 eV and 13,500 eV at a time resolution of 50 ms, but operated in a mode between 40 eV and 10,000 eV during the intervals studied in this article \citep{halekas2025tracers}. 

To contextualize our observations with respect to solar wind conditions, we use the OMNI \textit{OMNI\_HRO\_1MIN} data product at 1-minute time resolution \citep{papitashvili2014omni}. In OMNI, L1 observations of the pristine solar wind are from instruments on either the ACE \citep{mccomas1998solar} or WIND \citep{ogilvie1995swe} spacecraft, and are processed to account for propagation to Earth's bow shock. The OMNI data product includes the solar wind magnetic field vector (IMF $B_x$, $B_y$, and $B_z$ in geocentric solar magnetospheric coordinates [GSM]), as well as the solar wind plasma density, flow speed, and subsequent dynamic pressure. 

We also use measurements of the Cross-Polar Cap Potential (CPCP) from the Super Dual Auroral Radar Network \citep{greenwald1995darn,chisham2007decade} to compare against our reconnection rates. SuperDARN is a global collaboration of ground-based high-frequency radars that measure ionospheric convection via the line-of-sight velocity of field-aligned density irregularities. Line-of-sight measurements are taken from up to 30 Northern hemisphere radars. The data are gridded, combined, and then fitted with a series expansion of spherical harmonic functions to infer the electric potential of the polar cap \citep{ruohoniemi1998large}.  Example convection maps from the SuperDARN network for select events from the first storm studied (September 30, 2025) are shown in \ref{apx:convectionmaps} to demonstrate quality control. 

\section{Methodology} \label{sec:methodology}

\begin{figure}
    \centering
    \includegraphics[width=1\linewidth]{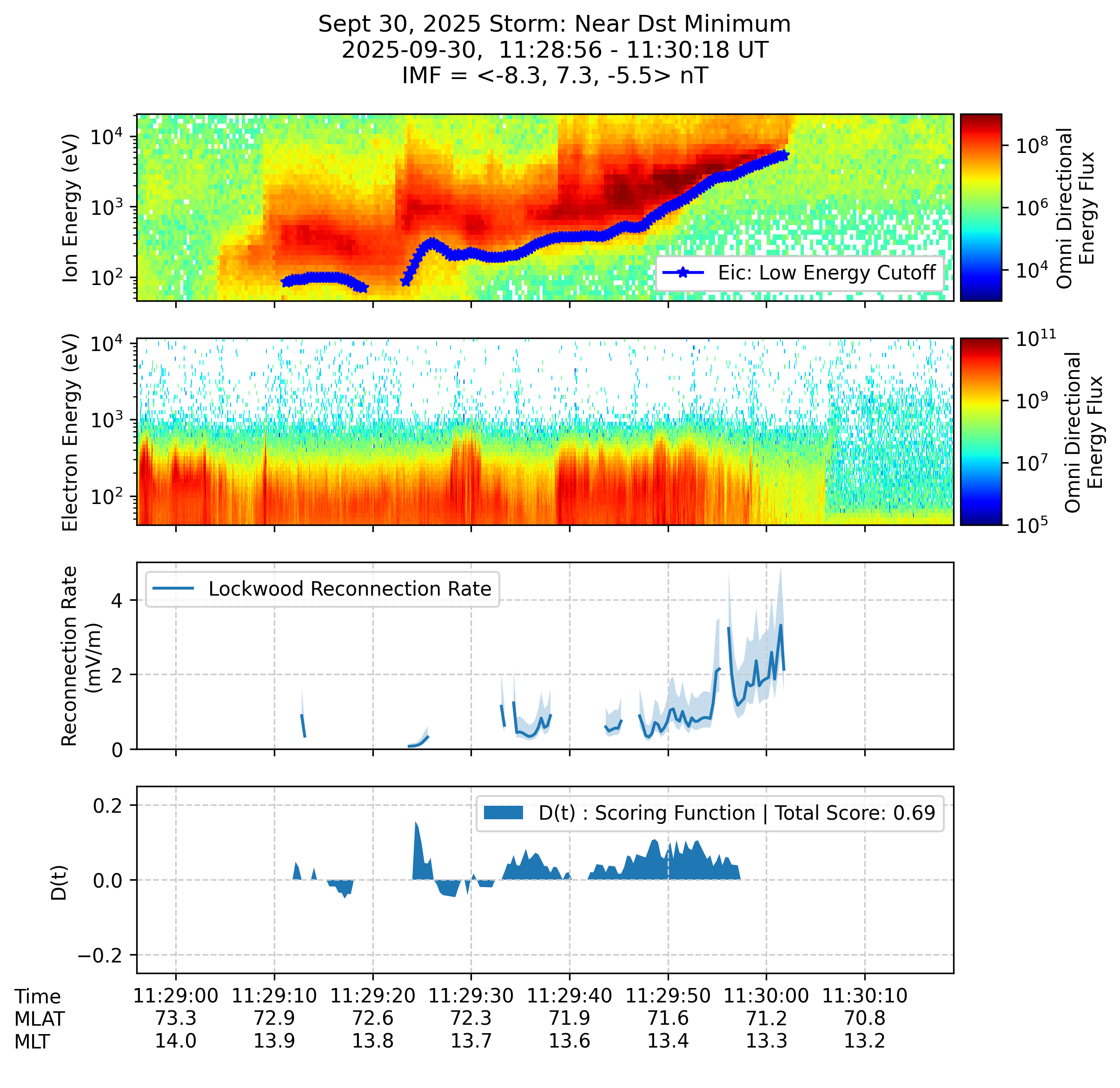}
    \caption{Dispersion event detected by TRACERS near the Dst minimum (-129.0 nT) of the September 30, 2025 storm. This event exhibits high fluxes with higher energy spread than during the recovery period, and is observed around 13.3-13.9 MLT.}
    \label{fig:event_sept30_dstmin}
\end{figure}

\subsection{Automatic Event Identification}
To collect events, we used previously developed algorithms to automatically identify ion-energy dispersion events, as described in \citet{da2022automatic}. The algorithms were previously developed with DMSP and are being applied to TRACERS for the first time here. While we use the algorithm to automatically identify events, we also manually review the data to ensure no events are missed. 

In the automatic identification algorithm, a lower energy cutoff $E_{ic}(t)$ is calculated at each timestep. This curve is defined as the energy at which the flux at that energy is 10\% of the peak flux across the entire ion energy spectrum \citep{lockwood1992variation}. An instantaneous scoring function $D(t)$ is applied to the data at each timestep based on $E_{ic}(t)$ and OMNI data, which determines a number of conditions that qualify it as an event which will be described shortly. To detect an event, a 60-second sliding window is moved across the data; if the total score within the window exceeds a given threshold, the algorithm determines that an event occurs.

\noindent The scoring function is,

\begin{equation} \label{eqn:scoring}
    D(t) = - \textrm{sgn}\left(\frac{d|MLAT|}{dt}\right) \left[ \frac{dLog [E_{ic}]_{smoothed}}{dt}\right] a(t) b(t) c(t) d(t) e(t),
\end{equation}

\noindent where $a(t)$, $b(t)$, $c(t)$, $d(t)$, and $e(t)$ are functions taking values of either 0 or 1, indicating whether a binary condition on the data holds. Flux units are defined here as the values obtained by taking the average of the ACI pitch angle distribution, yielding units of differential energy flux per unit steradian (eV cm$^{-2}$ s$^{-1}$ sr$^{-1}$ eV$^{-1}$). In the plots of this manuscript, these flux units are labeled  \textit{Omni Directional Energy Flux}. The conditions are as follows: $a(t) = 1$ when the average ion flux units between 50 eV and $3.1 \times 10^3$ eV (the nominal range of magnetosheath origin) exceeds $10^6$ flux units; $b(t) = 1$ when either electron data is unavailable or the average electron flux between  50 eV and $3.1 \times 10^3$ eV (the nominal range of magnetosheath origin) exceeds $10^7$ flux units summed over pitch angle; $c(t) = 1$ when the ion flux at $E_{ic}$ exceeds $10^{6.5}$ flux units; $d(t) = 1$ when $E_{ic}$ is under  $3.1 \times 10^3$ eV (the lower energy cutoff is nominally in the magnetosheath); and $e(t) = 1$ when the magnetic local time (MLT) is on the dayside. 

In this work, $E_{ic}(t)$ is smoothed with a moving window average with five points (1.56 seconds on each side for a total \ $\sim 3$ second window) which proved adequate to mitigate noise in the signal in this TRACERS/ACI data. We note that in Equation \ref{eqn:scoring}, $D(t)$ takes a positive score when the lower energy cutoff is changing in the direction associated with the dispersion phenomena. The final condition that determines selection is $S > \tau$, for score $S$ where,

\begin{eqnarray}
    S = \int_{window} D(t) dt.
\end{eqnarray}

An example of a selection and the associated $D(t)$ curve is found in Figure \ref{fig:event_sept30_dstmin}. The top panel shows the ion spectrogram, with $E_{ic}(t)$ as the blue curve. In the second panel is the electron spectrogram, in the third panel is the calculated reconnection rate (which will be addressed later), and in the bottom panel is the instantaneous scoring function $D(t)$. We notice that because of the binary functions ($a(t),..., e(t)$), the instantaneous score only takes positive values around the dispersion event, yielding a total score of 0.69. To select events for each storm in this study, we used $\tau=0.1$ and were able to automatically identify events previously found through a manual search.

\subsection{Calculation of the Dayside Magnetopause Reconnection Rate}

In this work we apply the methodology of \citet{lockwood1992variation} with enhancements made in \cite{da2025calculation} to estimate the reconnection rate from the lower energy cutoff of the dispersed ion population. In the classical 2D picture of ion-energy dispersion, the precipitating ions are injected along the dayside x-line, and become dispersed along convecting magnetic field lines due to time-of-flight (TOF) effects. In these TOF effects, lower-energy particles are given more time to drift across field lines than higher-energy particles, and in turn, the particles become spatially organized across the direction of the plasma convection. The approach of \citet{lockwood1992variation} is to relate the spread and structure of the dispersion to the upstream reconnection rate assuming that the structure is created entirely by temporal variability of reconnection at the magnetopause.

The reconnection rates calculated here are given as multiple values for each event, computed individually at each point along the lower-energy cutoff curve. The approach nominally uses the lower energy cutoff, $E_{ic}$, and derives the following equation to relate $E_{ic}$ and $dE_{ic}/dt$ to the reconnection rate $E_y'$:

\begin{equation} \label{eqn:recon_rate}
    E_y' = \left( \frac{dy}{dy'} \right) \frac{B_s V_s \textrm{cos}(\alpha)}{1 + (\frac{d'}{2})\sqrt{\frac{m}{2}} E_{ic}^{3/2} \frac{dE_{ic}}{dt}},
\end{equation}

\noindent where $B_s$ is the magnetic field strength at the spacecraft (taken from the dipole model due to lack of spacecraft measurements at this time), $V_s$ is the spacecraft velocity relative to Earth center, $\alpha$ is the angle between the spacecraft velocity and the direction of the field line convection (taken as $0^\circ$ due to absence of precise in-situ electric field measurements at these times), $d'$ is the length of the particle transit path adjusted for parallel electric field effects (neglected in this study), and $dy/dy'$ is a flux tube width scaling factor taken as $dy/dy' \approx \sqrt{B_{mp}/B_{s}}$ with $B_{mp}$, the magnetic field strength at the magnetopause, estimated as $50~nT$. 

Each of these field lines adds magnetic flux to the polar cap, and the rate at which they are added is related to the satellite velocity $V_s$ and the rate of newly opened field lines, related to $dE_{ic}/dt$. The rate at which magnetic flux added to the polar cap at the satellite location is estimated using $B_s$, and the induced electric field is calculated using Faraday's law.  This electric field is related to the upstream electric field using the mapping factor $dy/dy'$. One may wonder why the reconnection rate depends on $V_s$; this can be understood to be related to the fact that other variables such as $dE_{ic}/dt$ implicitly depend on the speed of the sampling spacecraft. 

An example of a dispersion event and its associated set of reconnection rates is shown in Figure \ref{fig:event_sept30_dstmin}. The top panel shows a dispersion structure with the associated $E_{ic}$ curve in blue. The second panel shows the electron spectrogram, displaying a long injection of magnetosheath electrons coinciding with the ion dispersion. The third panel shows the reconnection rates along the dispersion curve, which exhibit the typical behavior of producing a variable range with an increasing trend towards the high end of the dispersion structure.  The shaded error bars in the reconnection rate are calculated by repeating the calculation with $dE_{ic}/dt \rightarrow \frac{1}{2} dE_{ic}/dt$ and $dE_{ic}/dt \rightarrow \frac{3}{2} dE_{ic}/dt$ per the previous approach developed in \citet{da2025calculation}.

Major uncertainties arise from determining $E_{ic}$, $dE_{ic}/dt$, $\alpha$ , and $d'$.  Complications involved with the variables $E_{ic}$  and $dE_{ic}/dt$  arise from Poisson noise, energy resolution, and precipitating populations that don't fit a simple Maxwell-Boltzmann distribution. Variables like $\alpha$ and $d'$ are obtained through modeling, and the uncertainties in them relate to the model used and the input data driving it (such as OMNI). For a more in depth discussion of uncertainties in the calculation, the reader is referred to \citet{da2025calculation}.

\subsection{Maximum Magnetic Shear Model}
To estimate the particle transit distance $d'$, we incorporate modeling of the x-line location from the Maximum Magnetic Shear model (MaxMS; \citealp{trattner2017mms}). In the MaxMS model, the magnetosheath magnetic field is calculated using an analytic model based on L1 measurements and combined with the T96 magnetospheric magnetic field model \citep{tsyganenko1995modeling}. The x-line determined by the MaxMS model is based on the locations of magnetic shear between the magnetosphere and the magnetosheath.

To calculate $d'$, we combine the x-line output of the MaxMS model with field line traces using the T96 model. First, field line traces are started along each point of the MaxMS x-line within the neighborhood of the MLT of TRACERS. Due to T96 model uncertainties; especially in relation to the location of the observed cusp, the set of magnetospheric field lines may not trace to low altitude locations sufficiently near to TRACERS. In that case, the starting positions are adjusted inward (earthward) by 2.5\%, and the model field line traces are again determined. From these traces, we select the magnetic field line that maps closest to the satellite position. If the selected transit path maps to the edge (or outside) of the defined region (neighborhood), the process is restarted (iterated) using an expanded region. With the final transit path selected, the variable $d'$ is then the length of that path.

\subsection{Cusp Angle of Attack ($\alpha$)}

We use a model to calculate the $\alpha$ variable, which will appear later in our methodology section for calculating the reconnection rate. This variable, known as The angle of attack, represents the angle between the TRACERS spacecraft trajectory and a parameterized cusp model \citep{petrinec2025tandem}, expressed in the Altitude-adjusted Corrected Geomagnetic (AACGM) coordinate system. A small angle represents a direct traversal through the model cusp (essentially along a constant MLT); while a larger angle represents an oblique traversal through the model cusp. A mean angle of attack value is determined from the entry and exit cusp crossing points as the values may somewhat differ; especially for oblique traversals.

\section{Reconnection Rates During Two Storms} \label{sec:storms}

In this section, we study the September 30, 2025, and January 19, 2026, storms using TRACERS. The September 30, 2025 storm was a  storm with southward IMF and strong $B_y$ and $B_x$ (positive) components throughout the main phase. The January 19th, 2026 storm goes from a strongly northward IMF to a southward IMF during the main phase with a $B_y$ component but a small $B_x$ component. The first storm is believed by to be caused by both an interplanetary coronal mass ejection (ICME) and a co-rotating interaction region (CIR), while the second is caused only by an ICME. The causes by ICMEs are believed due to concurrent ICMEs being listed in the Richardson ICME catalog. The September 30, 2025 storm was previously studied in \citet{shirshunpublished}, which pointed to evolving cusp precipitation with a poleward shift, transition from standard to reverse dispersion, and increasing transit distances.

21 non-reverse ion-energy dispersion events were found during the September 30, 2025 storm, while only two were found during the January 19, 2026 storm. The events during the September 30, 2025 storm varied in MLT between 10 and 14 MLT, while the events from the January 19, 2026 storm occurred closer to dawn at 9.6-9.7 MLT. During the September 30, 2025 storm, TRACERS collected data in the northern cusp, while during the January 19, 2026 storm, TRACERS collected data in the southern cusp. It is expected that the reason for fewer events during the January 19, 2026, event is that TRACERS spent less time collecting data in the cusp, spending more time closer to dawn.

\begin{figure}
    \centering
    \includegraphics[width=1\linewidth]{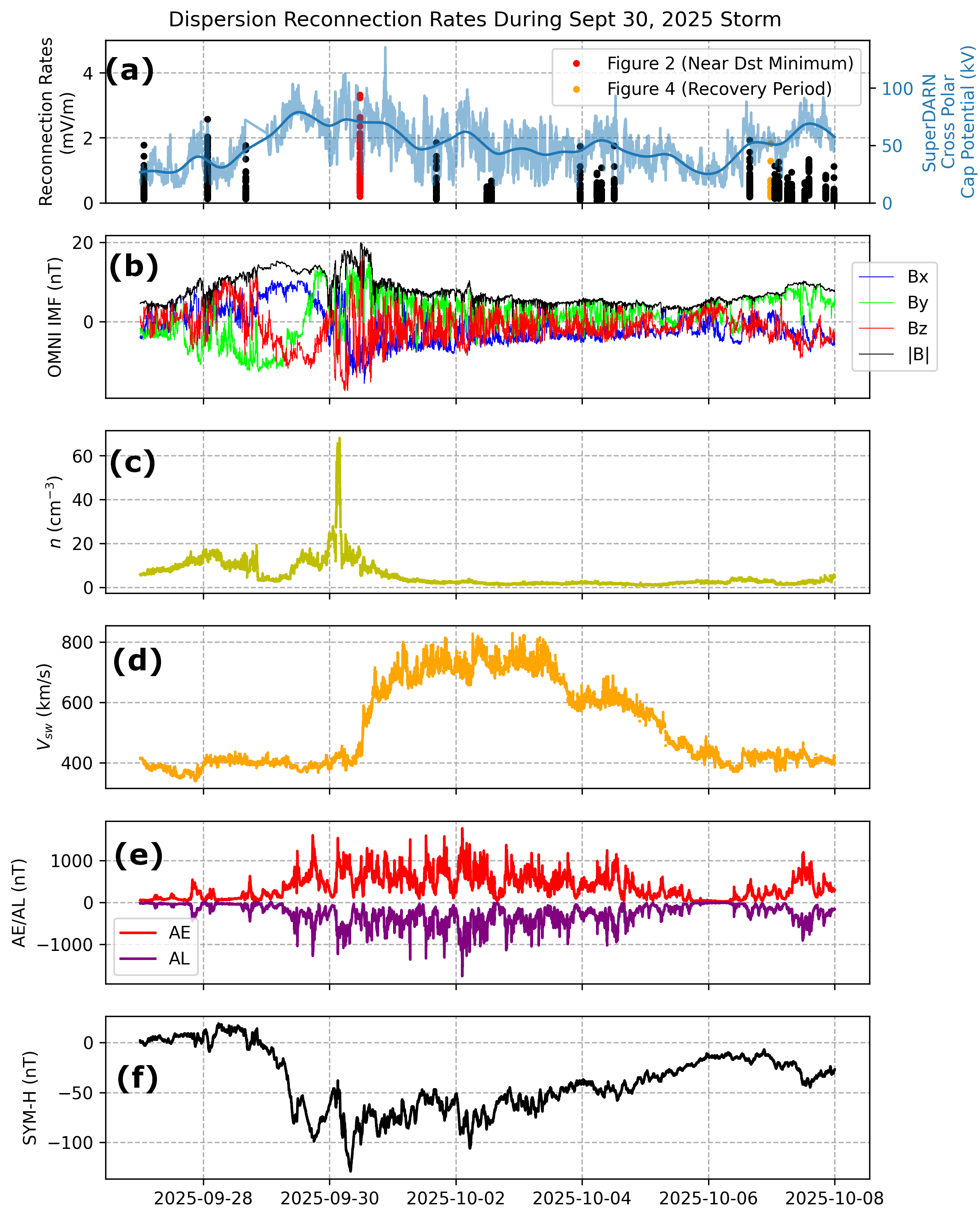}
    \caption{Reconnection Rate calculated from TRACERS cusp ion-energy dispersion during the September 30, 2025 storm, overlaid with Cross Polar Cap Potential (CPCP) from the SuperDARN radar network.}
    \label{fig:Sept30_Storm_recon_during_storm}
\end{figure}

\begin{figure}
    \centering
    \includegraphics[width=1\linewidth]{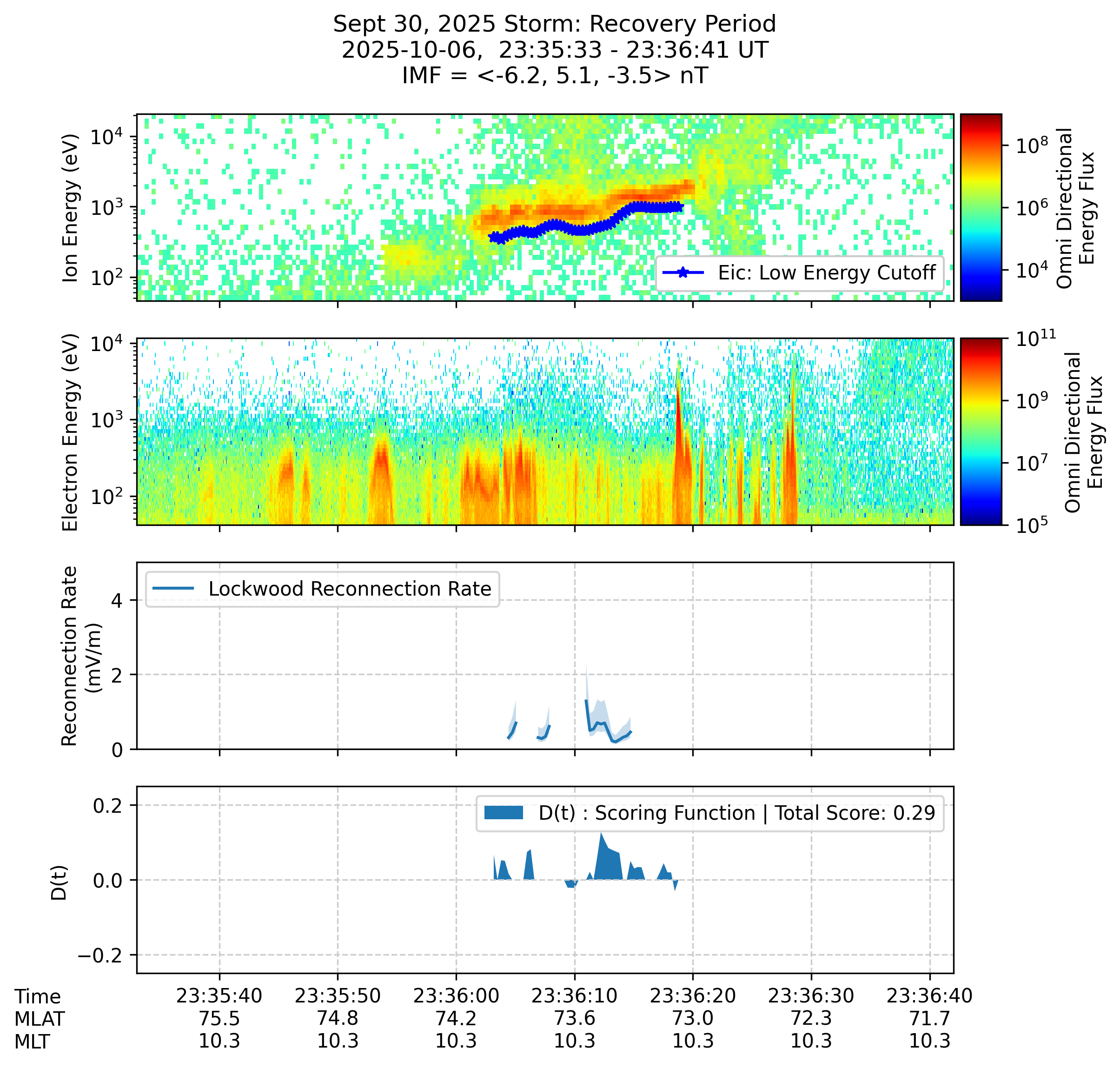}
    \caption{Dispersion event observed during the recovery period of the September 30, 2025 storm. It includes moderately high fluxes, and is observed around 10.3 MLT.}
    \label{fig:event_sept30_recovery}
\end{figure}

Reconnection rates were calculated for each of the dispersion events and compared to the cross polar cap potential (CPCP). We will first talk about the September 30, 2025, storm. Figure \ref{fig:Sept30_Storm_recon_during_storm} in panel (a) shows the reconnection rates for the 21 events during the storm, along with the CPCP during the same time period.  In Figure \ref{fig:Sept30_Storm_recon_during_storm} , panel (b)  shows the OMNI IMF conditions during the storm,  which contain a strongly negative $B_z$ up to 17.45 nT during the main phase of the storm, which returns closer to zero during the recovery phase. During the recovery phase are also  IMF fluctuations visible in all three components of the IMF.  Panel (c) shows the solar wind number density, which is low (around 10-20 cm$^{-3}$) up until around the time of the minimum SYM-H, where it briefly spikes to 68.1 cm$^{-3}$ before rapidly returning to low levels. Panel (d) shows the solar wind flow speed, which sits at around 400 km/s up until around the time of minimum SYM-H, where it becomes enhanced up to 829 km/s,  then gradually decreasing throughout the recovery period. The information in panels (c) and (d) indicate that there would be a rapid, sudden compression of the magnetosphere around the time of the spike in solar wind number density. 

The AE and AL indexes shown in panel (e), indicate substorm activity throughout the recovery period of the storm. The AE index (which is calculated as $AU-AL$ for a third related index AU) measures the strength of the auroral electrojets. Because substorms are dominated by a massive enhancement of the westward electrojet in the midnight sector, the most prominent substorm signature is driven by a sudden, deep drop in the AL index, which creates a massive spike in the AE index. Panel (f) shows the SYM-H index, a high time-resolution storm index comparable to Dst.  

The previously shown Figure \ref{fig:event_sept30_dstmin} details a dispersion event near the Dst minimum for the storm (-129 nT), and Figure \ref{fig:event_sept30_recovery} details a dispersion event during the recovery period of the storm.  

\begin{figure}
    \centering
    \includegraphics[width=1\linewidth]{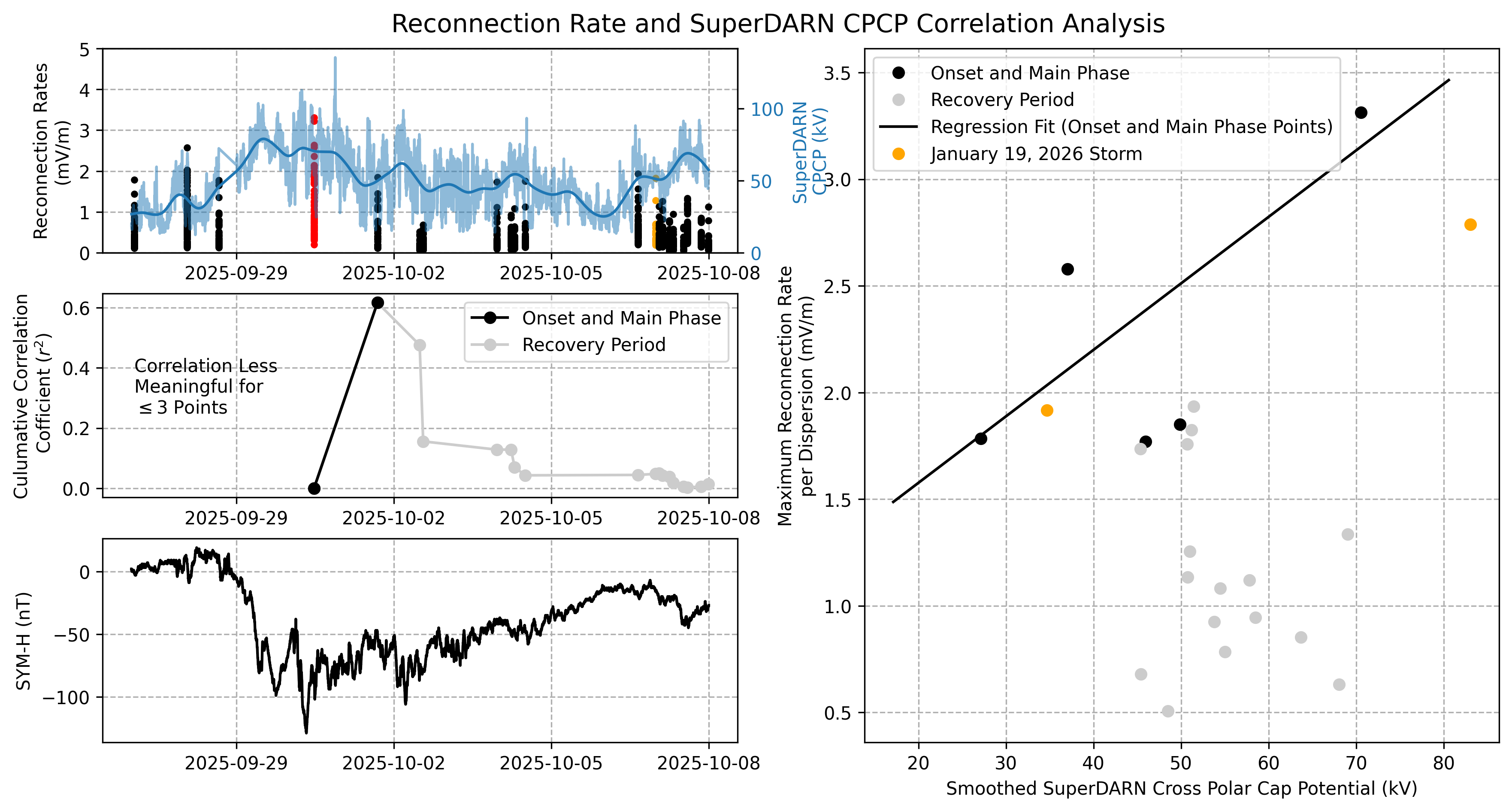}
    \caption{Correlation analysis between the maximum in-family reconnection rate per dispersion curve, and the smoothed CPCP. When including dispersion from the onset and main phase, a maximum correlation of $r^2=0.62$ was reached, which dropped off when including dispersions from the recovery period.}
    \label{fig:r2_analysis}
\end{figure}

In Figure \ref{fig:r2_analysis}, we observed that the reconnection rates track moderately well with the CPCP. We note that the CPCP is taken from the northern hemisphere, where the events are detected. For each dispersion curve, we selected the maximum reconnection rate calculated from the main reconnection rate curve. The selected values occur near the open/closed boundary. We purport that this is the best nominal reconnection rate for each curve, because it is related to the current magnetic flux being added to the polar cap; the values outside the open/closed boundary represent \textit{pending} flux to be added when that field line completes its convection. 

In quasistationary state, this rate of adding the open flux at the dayside should be balanced by removal of the open flux at the nightside. This is the basics of the Dungey cycle. In this scheme, the dayside reconnection potential will be approximately equal to the nightside reconnection potential and equal to CPCP. Substorms change the balance and could make deviations from the stationary CPCP making it larger or smaller. During the recovery phase of the storm, we expect that CPCP on average becomes smaller so deviations caused by substorms could destroy correlation between reconnection rate and CPCP. 

We note that while the CPCP and the reconnection rate are related, they are separate metrics tracking these two connected yet distinct quantities. The CPCP is inherently a global metric, representing an integral of the rate over the full reconnecting region. In contrast, the reconnection rate we calculate is localized to  a part of the x-line. For this reason, CPCP includes elements like the total spatial extent of reconnection or how the rate may vary along the reconnecting region, which is absent from our localized reconnection rate.

In the top left panel of Figure \ref{fig:r2_analysis}, we show the reconnection rates for each dispersion curve along with the CPCP and a smoothed version of the CPCP (smoothed using a smoothing spline with $\lambda=0.25$, \citealp{virtanen2020scipy}). Correlations were prepared at different cutoff points in time between the\ maximum value of each dispersion curve, which occurs near the open/closed boundary in each case, and the smoothed CPCP. In the middle left panel, the correlation coefficient $r^2$ is shown between these two sets at increasing cutoff times. The time of the best cutoff is after five dispersions, at October 2, 2025, 11:47:46 UT, which coincides with the end of the main phase of the storm. When the recovery time was included in the correlation, the outcome $r^2$ successively drops to near zero.  In the right panel of Figure \ref{fig:r2_analysis}, a scatterplot with the line of best fit is presented with the points grouped between those in the onset and main phase, and those in the recovery period. 

The correlation between the reconnection rate from dispersion and the CPCP may be disrupted by multiple mechanisms.  During the main phase of the storm, the reconnection rate obtained from cusp dispersion is associated with dayside reconnection and addition of magnetic flux to the polar cap on the dayside \citep{cowley1992excitation}. During recovery time, flux is gradually removed from the polar cap through tail reconnection, which is not addressed by the reconnection rate calculation in this work. The effects of tail reconnection would not be expected to be correlated.




A second phenomenon which could contribute to breaking of the correlation is substorm activity. Furthermore, this storm exhibits enhanced substorm activity during the recovery period, which would contribute to the correlation, as the  loading and unloading of solar wind energy occurs on substorm time scales. Evidence of the substorm activity can be seen in panel (e), showing the AE and AL indeces \citep{forsyth2015new}. Moveover, while the local measurements from TRACERS are highly sensitive to the fine-scale temporal variations caused by the IMF fluctuations, the CPCP represents a macroscopic, integrated magnetospheric response. Because the magnetosphere-ionosphere system acts as a low-pass filter, the global CPCP naturally smooths out these local, high-frequency variations, further reducing the direct correlation between the two metrics.   The two orange points in Figure \ref{fig:r2_analysis} show the reconnection rates obtained from the January 19, 2026 storm for comparison. They have similar qualities to those found to be in correlation in the first storm. In the January 19, 2026 storm, these two points are spaced between the main phase and recovery period. A notable difference between that January 19, 2026 storm and the September 30, 2025 storm is that the September storm exhibits substorm activity during the recovery phase, whereas the substorm activity between the two points in the January 19, 2026 storm is less pronounced.

In this line of thoughts we assumed that dayside x-line length remains constant, but this assumption is not necessarily true. If it is the x-line length that dictates the reconnection potential at the dayside (with reconnection rate being approximately constant and not exceeding 3 mV/m) our results may suggest that during the recovery phase of storms when reconnection rate is smaller than during the main phase, the reconnection potential is governed mainly by the length of x-line (measured by CPCP).

In the individual events of Figure \ref{fig:event_sept30_dstmin} (near Dst minimum) and Figure \ref{fig:event_sept30_recovery} (recovery time), the event near Dst minimum showed a much hotter dispersed population with much higher fluxes than during the recovery period. Furthermore, the dispersion extends to higher energies in the Dst minimum event than in the recovery time event. An interpretation of these differences is that precipitation is more intense during the main phase of the storm, due to higher dynamic pressure and more intense ongoing reconnection.   We note that electron spectrogram data were available for the Dst minimum event but not the recovery period event.

\begin{figure}
    \centering
    \includegraphics[width=1\linewidth]{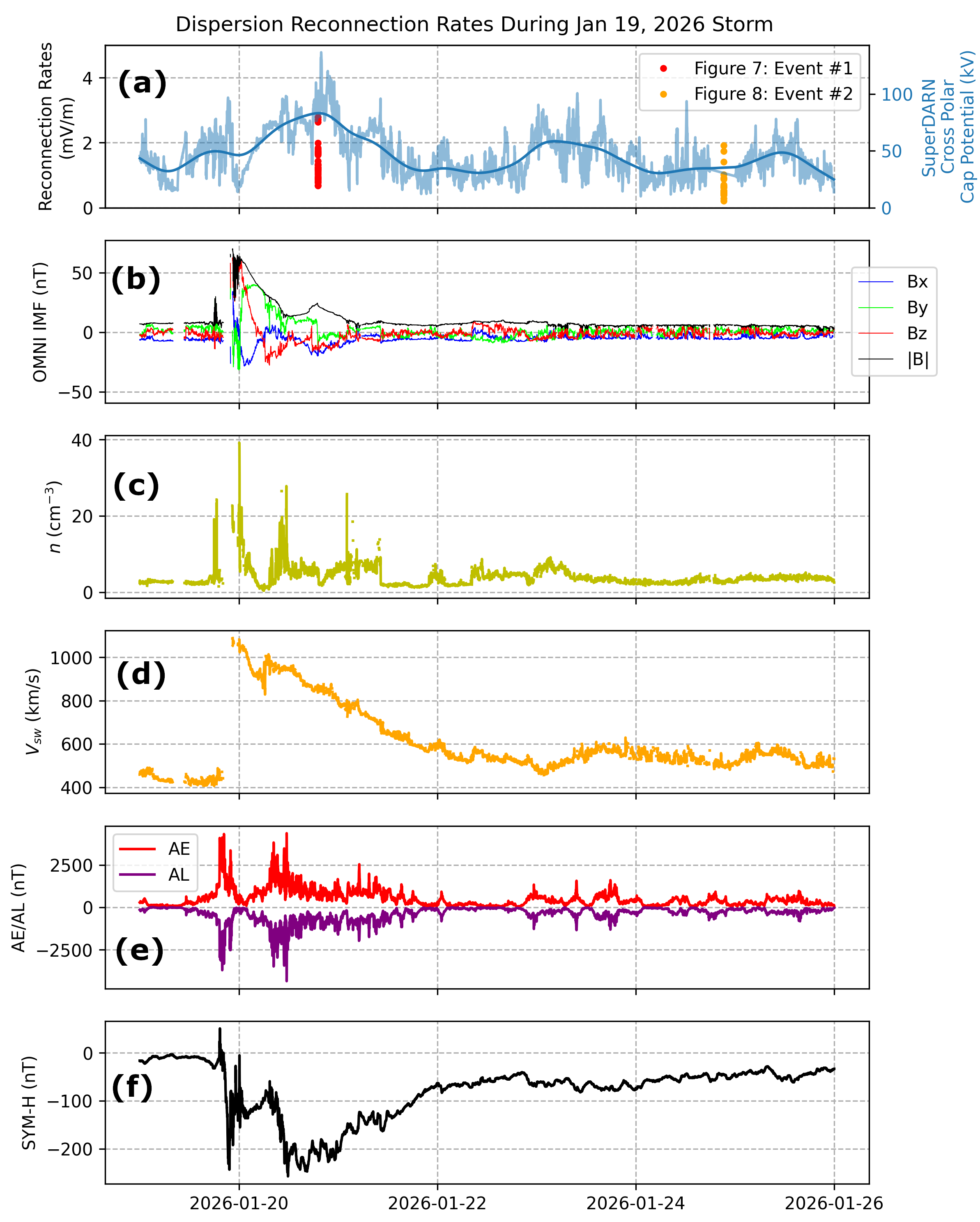}
    \caption{Reconnection Rate calculated from TRACERS cusp ion-energy dispersion during the January 19, 2026 storm, overlaid with Cross Polar Cap Potential (CPCP) from the SuperDARN radar network.}
    \label{fig:Jan19_Storm_recon_during_storm}
\end{figure}

\begin{figure}
    \centering
    \includegraphics[width=1\linewidth]{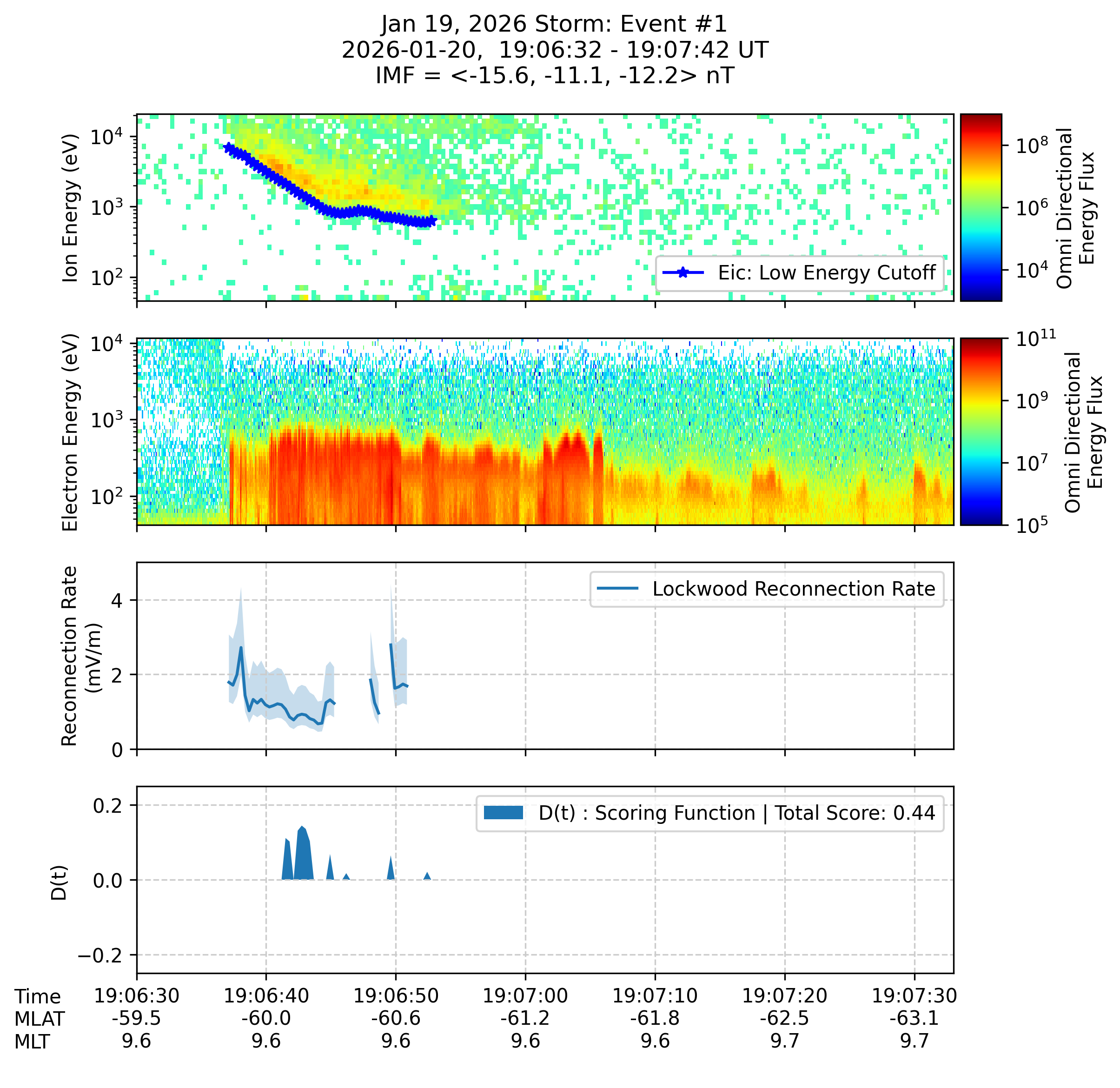}
    \caption{First dispersion event identified during the January 19, 2026 storm. This event was observed around 9.6 MLT, which may be dawnward outside the cusp, and includes relatively low ion fluxes.}
    \label{fig:event_jan19_event0}
\end{figure}

\begin{figure}
    \centering
    \includegraphics[width=1\linewidth]{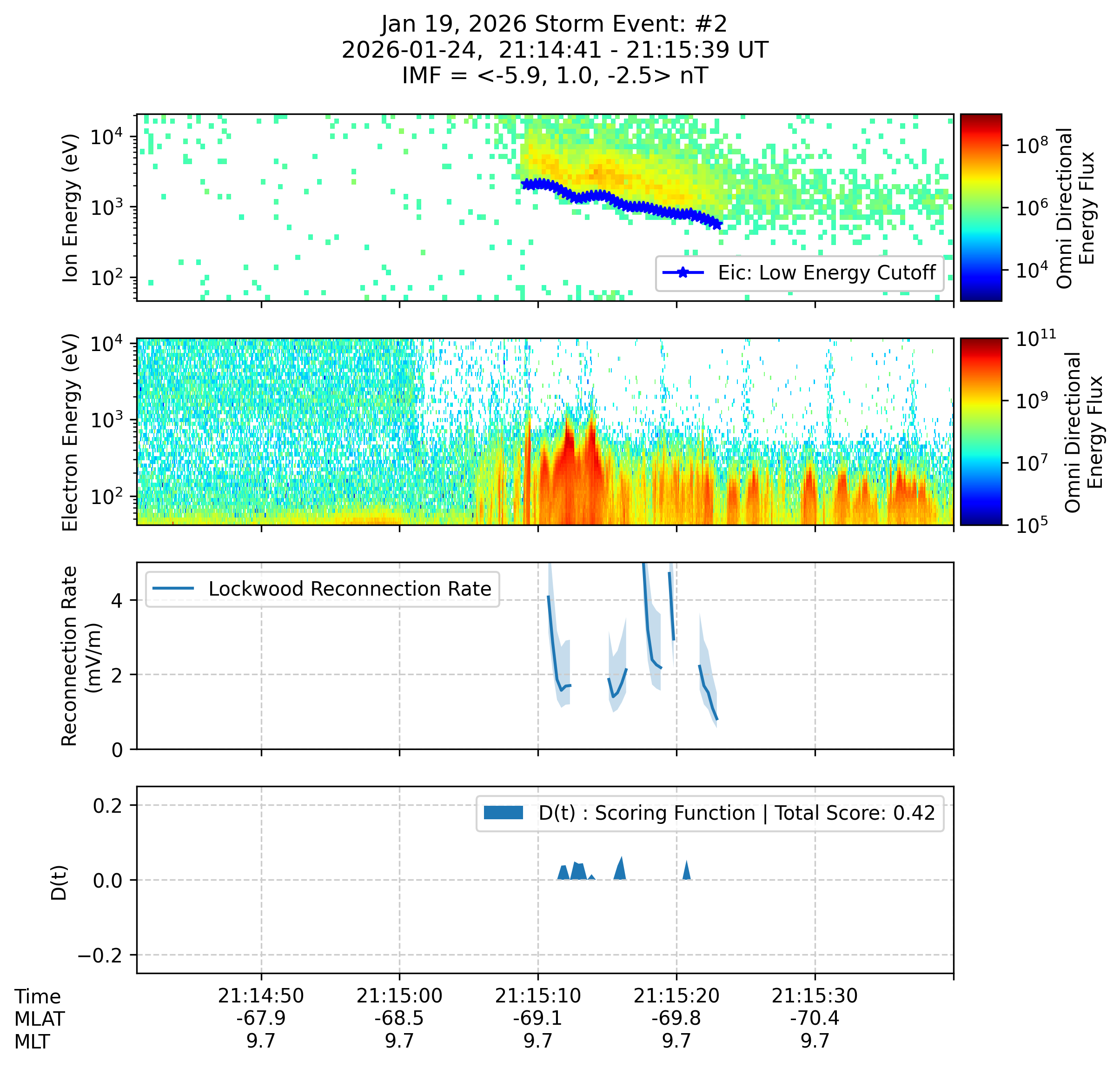}
    \caption{Second dispersion event identified during the January 19, 2026 storm. This event was observed around 9.7 MLT, which may be outside the cusp, and also includes relatively low ion fluxes.}
    \label{fig:event_jan19_event1}
\end{figure}

In Figure \ref{fig:Jan19_Storm_recon_during_storm}, Panel (a) shows the reconnection rates and CPCP for the January 19th, 2026 storm, with the panel layout similar to Figure \ref{fig:Sept30_Storm_recon_during_storm}.  During the start of the storm, $B_z$ is northward, and turns southward coinciding with the second dip in the SYM-H index (Panel (f)). In panel (c), we see several isolated enhancements in the number density that occur throughout the main phase of the storm. Panel (d) shows the solar wind flow speed, which jumps to over 1000 km/s during the main phase of the storm and gradually decreases back to under 600 km/s. 

We note that for this storm, the dispersion events were detected in the southern cusp, but the CPCP is from the northern hemisphere because SuperDARN measures the northern-hemisphere CPCP much more reliably than in the southern hemisphere. 

During the January 19, 2026, storm, only two dispersion events were detected.  Figures \ref{fig:event_jan19_event0} and \ref{fig:event_jan19_event1} show these two events. They were cold (low energy spread), low-to-moderate flux dispersion events detected at 9.6 MLT and 9.7 MLT, respectively.  During both of these events, $B_z$ was southward. In this case, the first event was close to the Dst minimum for this storm, and the second event was close to the recovery period, but the Dst minimum event was not hotter and of higher flux as in the September 30, 2025 storm.

When comparing the reconnection rate obtained from dispersion to the CPCP, care must be taken. Reconnection rates obtained from dispersion correspond to a localized piece of the x-line, which may exhibit variation in the reconnection rate across MLT. The CPCP, on the other hand, is a global metric. Still, the reconnection rates in Figure \ref{fig:Sept30_Storm_recon_during_storm} show that the higher ends of the reconnection rates collected for each event tend to correspond to the peaks in the CPCP throughout the storm, suggesting the ability to accurately quantify the global reconnection process despite measuring a localized piece of the x-line.

\begin{figure}
    \centering
    \includegraphics[width=1\linewidth]{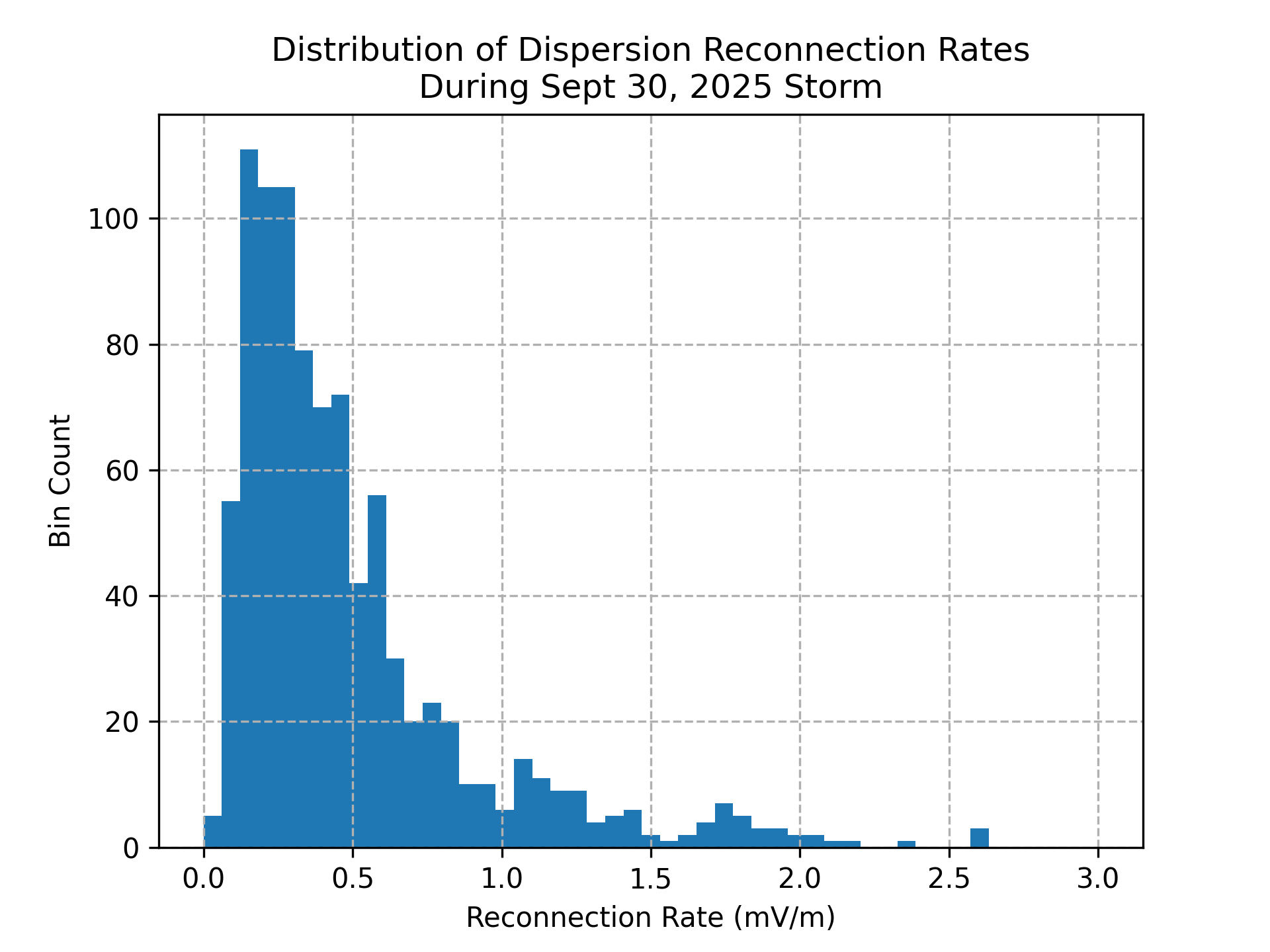}
    \caption{Histogram of reconnection rates calculated throughout the September 30, 2025, storm using TRACERS. }
    \label{fig:dist_of_recon_rates}
\end{figure}

We also present the distribution of reconnection rates throughout the September 30, 2025, storm in Figure \ref{fig:dist_of_recon_rates}. The reconnection rates peak around 0.25 mV/m and span about 0.1-1.5 mV/m. This range is slightly less than the range obtained from cusp dispersion with DMSP in \citet{da2025calculation}, which was 0.1 - 2 mV/m. It is however similar to in situ measurements of the reconnection rate, such as \citet{mozer2007quantitative}, where an average reconnection rate of 1.25 mV/m was obtained from eleven reconnection events using observations from the Polar mission. In the MMS era, a mean magnetopause reconnection rate of 1.164 mV/m with a standard deviation of 0.376 was found  \citep{burch2020electron}.  To compare to the magnetotail, the work of \citet{bessho2018effect} estimates a much stronger reconnection rate of 4-5 mV/m by indirect comparison of modeled  crescent distribution functions. 

\section{Discussion and Conclusion} \label{sec:conclusion}

In our first look at TRACERS data, we see that TRACERS observes cusp dispersion on the dayside, similar to DMSP, but at much higher temporal resolution. This higher time resolution translates to better estimates for $E_{ic}$ and $dE_{ic}/dt$, the two key time-varying parameters used in the reconnection rate calculation (Equation \ref{eqn:recon_rate}). In addition to the improved temporal resolution, the instrumentation aboard TRACERS is free of FOV-bias that affected DMSP, wherein only 90$^\circ$ of the pitch angle distribution was measured by DMSP at any given time. With the increased time resolution of TRACERS/ACI, we were able to obtain more reconnection rate calculations per dispersion event than DMSP.

We provided a correlation analysis between the dayside reconnection rate calculated from cusp dispersion and the CPCP, and identified a strong correlation of $r^2=0.62$ during the onset and main phase of the storm. We identified this correlation falls off during the recovery period, which is to be expected from knowledge of global dayside versus nightside competition to influence the CPCP throughout geomagnetic storms.

We also applied and validated automatic event identification to the TRACERS data, detecting all events we had found by hand. This has the potential to be a useful tool later in the TRACERS mission for statistical analysis like \citet{da2024statistical}, when the volume of data exceeds the capacity to manually select dispersion events. 

To generate these first results, we used $\alpha=0^\circ$ in the reconnection rate calculation found in Equation \ref{eqn:recon_rate}. This variable can, in principle, be estimated using in-situ electric field measurements \citep{da2025calculation}. This was not done in this manuscript due to mission challenges that prevented precise attitude solutions and subsequently accurate electric field measurements. However, the issue is being actively addressed and is expected to be corrected in the future. 

The peak reconnection rates we calculated tracked fairly well with the CPCP from SuperDARN, despite reconnection rates being a more localized measure of an x-line, whereas CPCP is a global measure.  Given the agreement shown here and the ability to calculate reconnection rates multiple times per geomagnetic storm, with further development, reconnection rates derived from ion-energy dispersion have the potential to serve as another metric of geomagnetic activity.

\appendix
\section{Example SuperDARN CPCP Convection Maps for Select
Events in the September 30, 2025 Storm} \label{apx:convectionmaps}

In this section, we display some diagnostic plots from the SuperDARN radar network to illustrate good coverage during the September 30, 2025 storm. Because data is captured approximately every 2 minutes, an exhaustive demonstration of sufficient coverage is outside the scope of this article, but the plots are made available online at \url{https://superdarn.ca/convection-maps}.

\begin{figure}
    \centering
    \includegraphics[width=1\linewidth]{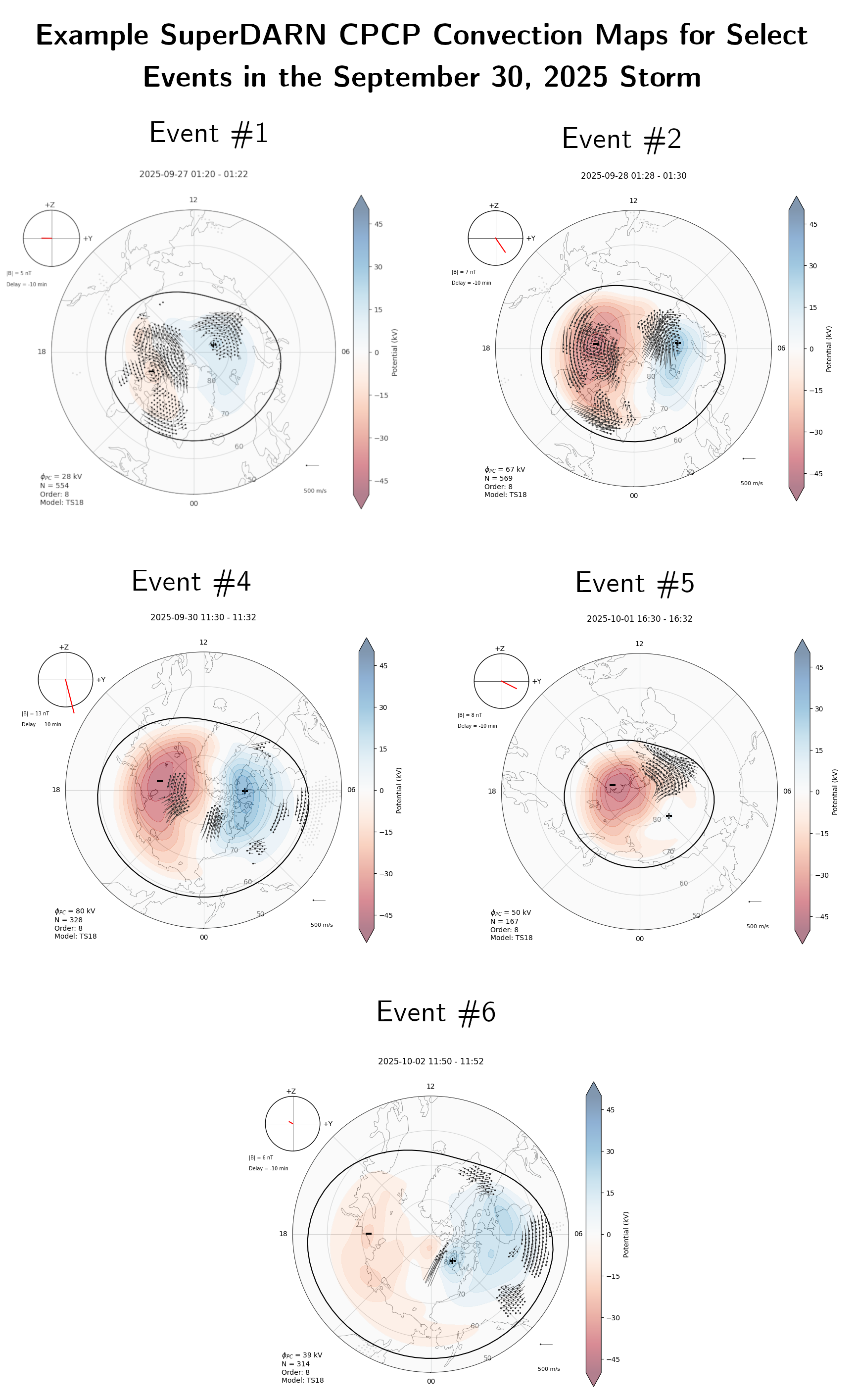}
    \caption{Example convection maps with accompanying data processing statistics ($N$) for select events from the September 30, 2025 storm. There is no convection map available for Event \#3 due to a failed spherical harmonic fit; in this manuscript, the CPCP at this time is interpolated over.}
    \label{fig:convectionmaps}
\end{figure}

Figure \ref{fig:convectionmaps} shows SuperDARN convection maps for the first six events of the September 30, 2025 storm, except event \#3. Event \#3 is absent due to spherical harmonic processing failing at that timestep. In this manuscript, we interpolate the CPCP at this time. The $N$ value in the bottom left of each plot indicates the number of points that contributed to each solution. A sufficient, but not necessary, condition for a good solution is $N>200$ \citep{walach2022super}, which holds for all events except event \#5. In event \#5, $N=167$, but this is retained because preceding solutions captured sufficient statistics and this solution does not cause a drastic variation in the CPCP.

%



%
%

\section*{Open Research Section}
TRACERS, OMNI, and SuperDARN CPCP data used in this study will be uploaded to Zenodo upon publication. The automatic identification code, as well as the code used to calculate the reconnection rate and to generate the plots in this manuscript, will also be uploaded to Zenodo. Updates to the reconnection rate code after publication will be made available online at \url{https://github.com/ddasilva/tracers-first-results-paper}.

TRACERS/ACI and TRACERS/ACE Level 2 data can be obtained at \url{https://tracers.physics.uiowa.edu/l2-public-data-products}.

SuperDARN RAWACF data is available with DOI from the FRDR repository \url{https://www.frdr-dfdr.ca/repo/collection/superdarn}. SuperDARN  convection maps are available from the SuperDARN team at \url{https://superdarn.ca/convection-maps}.
Line-of-sight velocities are calculated from RAWACF SuperDARN data using the FITACF 3.0 algorithm. The data is despecked before gridding, and then processed into convection maps describing the electric potential of the Northern polar cap. OMNI data of the interplanetary magnetic field (IMF) is supplied to the processing pipeline to supplement the ionospheric convection data with an empirical model based on statistical dependence of convection to the IMF \citep{thomas2018statistical}. The OMNI data were obtained from the GSFC/SPDF OMNIWeb interface at \url{https://omniweb.gsfc.nasa.gov}.
CPCP is calculated as the difference between the maximum and minimum potential estimated over the polar cap. The integration time for each map record that contains CPCP is 2 minutes.
All processing of SuperDARN data was done with the Radar Software Toolkit v5.1 (SuperDARN Data Analysis Working Group, et al. 2025). The documentation describes the process for gridding and making convection maps \url{https://radar-software-toolkit-rst.readthedocs.io/en/latest/}.

\section{Acknowledgments}
The authors thank the TRACERS team for the hard work and dedication to enable the TRACERS mission to come to fruition.

The authors acknowledge the use of SuperDARN data. SuperDARN is a collection of radars funded by national scientific funding agencies of Australia, Canada, China, France, Italy, Japan, Norway, South Africa, United Kingdom, and the United States of America.
The authors acknowledge the use of the Radar Software Toolkit version 5.1 for processing SuperDARN data, \url{https://doi.org/10.5281/zenodo.4435150}.

Co-author Gonzalo Cucho-Padin was supported by the NASA Living with a Star (LWS) program under grant 80NSSC23K0896.

\section*{Conflict of Interest declaration}
\noindent The authors declare there are no conflicts of interest for this manuscript.


%
\bibliography{references} 
%


%
%
%
%
%

\end{document}